\documentstyle[twoside,fleqn,espcrc2,epsf]{article}
\def\siml{{\ \lower-1.2pt\vbox{\hbox{\rlap{$<$}\lower6pt\vbox{\hbox{$\sim$}}}}\ }}

\def\lQ{\Lambda_{\rm QCD}}

\begin{document}
\newcommand{\ttbs}{\char'134}
\newcommand{\AmS}{{\protect\the\textfont2
  A\kern-.1667em\lower.5ex\hbox{M}\kern-.125emS}}

\hyphenation{author another created financial paper re-commend-ed}

\title{Heavy Hybrids in QCD effective theories}

\author{Nora Brambilla\address{Institut f\"ur Theoretische Physik,
        Boltzmanngasse 5, \\
        A-1090 Vienna, Austria}%
        \thanks{Marie Curie fellowship, Contract No. ERBFMBICT961714; 
                'Acciones Integradas' 1999-2000, project No. 13/99.}}

\begin{abstract}
We present model independent predictions on the short range behavior of the 
energies of the gluonic excitations between static quarks (hybrid
potentials) in an effective field theory framework (pNRQCD). 
\end{abstract}

\maketitle

\section{INTRODUCTION}
One of the clearest way of studying the heavy hybrid systems is from the energies of the 
gluonic excitations between static quarks, the so called excited gluonic potentials $V_H$.
These energies  are calculated on the lattice using Wilson loops with insertions of operators 
of a definite symmetry (see below)  in the end point strings. The frame is supplied by the 
Born-Oppenheimer approximation. In this picture the quarks and the 
gluons are respectively slow and fast degrees 
of freedom and in the limit ${\lQ/m}\to \infty$, ($m=$ mass of the quark), the static gluonic energies
 $V_H$ are treated as adiabatic potentials for the $Q\bar{Q}$ pair and   
inserted in the Schr\"odinger equation to get the masses of the hybrids.
Of course, the shape of $V_H$ as a function of the $Q\bar{Q}$ distance
 $r$  is   an interesting observable 
 of QCD.\par
Here, we  show how the appropriate  QCD effective field theory description of the heavy 
quark bound systems can  provide us with model independent predictions on the short range 
behavior of $V_H$.

\section{pNRQCD}
The existence of a hierarchy of energy scales $m\gg mv\gg mv^2$, $v\ll 1 $ being the  quark velocity,
 in the  heavy quark non-relativistic bound systems 
allows us to substitute scale by scale QCD with simpler but equivalent effective theories.
Integrating out the hard scale $m$ one passes from QCD to NonRelativisticQCD (NRQCD), while 
integrating out the soft scale $mv$  one passes from NRQCD to potential NRQCD (pNRQCD) \cite{pnrqcd}.  
In this last effective theory  only ultrasoft (US=with energy  much smaller than 
$mv$) degrees of freedom remain 
dynamical. The surviving fields are the $Q$-$\bar Q$ states  (with US energy) 
and the US gluons.  The $Q$-$\bar Q$ states can be decomposed into a singlet (S) 
and an octet (O) under color transformation. The relative coordinate ${\bf r}= {\bf x}_1-{\bf x}_2$, 
whose typical size is the inverse of the soft scale, is explicit and can be considered as small 
with respect to the remaining (US) dynamical lengths in the system. Hence the gluon fields
can be systematically expanded in $\bf r$ (multipole expansion). Therefore the pNRQCD Lagrangian 
is constructed not only order by order in $1/m$, but also order by order in 
${\bf r}$. As a typical feature of an effective theory, all the non-analytic 
behavior in ${\bf r}$ is encoded in the matching coefficients, which can be interpreted as potential-like terms. 
The  pNRQCD Lagrangian density is given at the next to  leading order in the multipole 
expansion by:
\begin{eqnarray}
& & \!\!\!\!\!\!\!\!\!\!\!\!\!\! {\cal L}_{\rm pNRQCD} =
{\rm Tr} \Biggl\{ {\rm S}^\dagger \left( i\partial_0 - {{\bf p}^2\over m} 
- V_s(r) + \dots  \right) {\rm S} 
\nonumber \\
& & \!\!\!\!\!\!\!\!\!\!\!\!\!\! \qquad + {\rm O}^\dagger \left( iD_0 - {{\bf p}^2\over m} 
- V_o(r) + \dots  \right) {\rm O} \Biggr\}
\nonumber\\
& & \!\!\!\!\!\!\!\!\!\!\!\!\!\! \qquad + g V_A ( r) {\rm Tr} \left\{  {\rm O}^\dagger {\bf r} \cdot {\bf E} \,{\rm S}
+ {\rm S}^\dagger {\bf r} \cdot {\bf E} \,{\rm O} \right\} 
\nonumber \\
& & \!\!\!\!\!\!\!\!\!\!\!\!\!\! \qquad  + g {V_B (r) \over 2} {\rm Tr} \left\{  {\rm O}^\dagger {\bf r} \cdot {\bf E} \, {\rm O} 
+ {\rm O}^\dagger {\rm O} {\bf r} \cdot {\bf E}  \right\},  
\label{pnrqcd0}
\end{eqnarray}
where ${\bf R} \equiv ({\bf x}_1+{\bf x}_2)/2$, ${\rm S} = {\rm S}({\bf r},{\bf R},t)$ and 
${\rm O} = {\rm O}({\bf r},{\bf R},t)$ are the singlet and octet wave functions respectively. All the gauge fields in Eq. (\ref {pnrqcd0}) are evaluated 
in ${\bf R}$ and $t$.  In particular ${\bf E} \equiv {\bf E}({\bf R},t)$ and 
$iD_0 {\rm O} \equiv i \partial_0 {\rm O} - g [A_0({\bf R},t),{\rm O}]$. 
$V_s$ and $V_o$ are the singlet and octet matching potentials;
$V_A$ and $V_B$ are matching coefficients. \par
In QCD another physically relevant scale has to be considered, i.e $\Lambda_{QCD}$, the 
scale at which the nonperturbative effects become relevant. As far as $mv\gg \Lambda_{QCD}$,
the matching between NRQCD and pNRQCD can be performed perturbatively.\par
We believe 
that pNRQCD is the most appropriate effective theory to describe the physics of heavy quark 
bound systems. For details see \cite{pnrqcd}.

\section{GLUELUMPS in pNRQCD\cite{pnrqcd}}
Although pNRQCD is originally designed to study $Q$-$\bar Q$ systems of large but finite mass, 
it is interesting to study its static limit.
We shall restrict ourselves to the case of pure gluodynamics.
We call gluelump the adjoint source in presence of a gluonic field 
\begin{equation}
H({\bf R},{\bf r},t) = H^a({\bf R},t) O^a({\bf R}, {\bf r}, t).
\label{gluelump}
\end{equation}
In the static limit there are no space derivatives in the Lagrangian, and hence {\bf r} and {\bf R} are good 
quantum numbers. The spectrum then consists of static energies which depend on {\bf r} 
(translation invariance forbids {\bf R} dependences) and on the only other scale in this problem, $\lQ$. 

In the limit $\lQ \ll 1/r$, the spectrum of the theory can be read from  
the Lagrangian (\ref{pnrqcd0}). In particular, the leading order solution corresponds to the zeroth order
of the multipole expansion. At this order the dynamics of the singlet and octet
fields decouple. Hence, the gluonic excitations between static 
quarks in the short-distance limit correspond to the gluelumps. Depending on the  glue operator $H$ and 
its symmetries, the gluelump operator  (\ref{gluelump}) describes a specific gluonic 
excitation between static quarks and its static energy $V_H$.

In NRQCD (as in QCD) the  gluonic excitations between static quarks have the same symmetries 
of a diatomic molecule plus charge conjugation. 
In the centre-of-mass system these correspond to the symmetry group $D_{\infty
  h}$ (substituting the parity generator by CP). 
According to it the mass eigenstates are classified in terms of the angular momentum 
along the quark-antiquark axes ($|L_z| = 0,1,2, \dots$ to which one gives the traditional 
names $\Sigma, \Pi, \Delta, \dots$), CP (even, $g$, or odd, $u$), and the
reflection properties  with respect to 
a plane passing through the quark-antiquark axes (even, $+$, or odd, $-$). 
Only the $\Sigma$ states are not degenerate with respect to the reflection symmetry.
In Fig. 1 we report the lattice measurement of $V_H$ obtained considering Wilson loops 
with operators of the appropriate symmetry inserted at the end points 
 \cite{Morningstar,Michael}. \par
In pNRQCD at lowest order in the multipole expansion, besides the already 
mentioned symmetries, extra symmetries for the gluonic excitation between static quarks appear.  The glue 
dynamics no longer involves   the relative coordinate $\bf r$. Therefore, 
the glue associated with a gluonic excitation between static quarks acquires a spherical symmetry.
In the centre-of-mass system gluonic excitations between static quarks are, therefore, classified 
according to representations of $O(3) \times$ C, which we summarize by $L$, the angular momentum, CP 
and reflection with respect a plane passing through the quark-antiquark axes.
 Since this group is larger than that one of NRQCD, several 
gluonic excitations between static quarks are expected to be approximately {\it degenerate} in pNRQCD, 
i.e. in the short-distance limit
$r \ll 1/\lQ$.
We illustrate this point by building up all operators, $H$, up to 
dimension 3 and by classifying them according to their quantum numbers in NRQCD and pNRQCD in Tab. \ref{tab3}. 
In Tab. \ref{tab3} all the operators are intended evaluated in the centre-of-mass coordinates. 
$\Sigma_g^+$ is not displayed since it corresponds to the singlet state. 
The prime indicates excited states of the same quantum numbers. 
The chosen operators for the $\Pi$ and $\Delta$ states are not eigenstates of the reflection operator. 
This is not important since these states are degenerate with respect to this symmetry.
The operators ${\bf E}$ and ${\bf D}\times {\bf B}$ (${\bf B}$ and ${\bf D}\times {\bf E}$) 
have the same quantum numbers and are
related by the equations of motion so that they  project over the same states. 
From the results of Tab. \ref{tab3} the following degeneracies are expected in
the short-distance limit:
\begin{eqnarray}
&&\Sigma_g^{+\, \prime} \sim \Pi_g\;; \qquad
\Sigma_g^{-} \sim \Pi_g^{\prime} \sim \Delta_g\,; 
\nonumber
\\
&&
\Sigma_u^{-} \sim \Pi_u\;; \qquad
\Sigma_u^{+} \sim \Pi_u^{\prime} \sim \Delta_u \,.
\label{dege}
\end{eqnarray}  
Similar observations have also been made in \cite{Foster}. In pNRQCD they  emerge  in a quite clear 
and straightforward way. Moreover, here we can write explicitly the relevant operators.
\begin{table*}[hbt]
\setlength{\tabcolsep}{1.5pc}
\newlength{\digitwidth} \settowidth{\digitwidth}{\rm 0}
\catcode`?=\active \def?{\kern\digitwidth}
\caption{ Matching operators $H$ for the $\Sigma$, $\Pi$ and $\Delta$ gluonic excitations between static quarks in 
pNRQCD up to dimensions 3.
 The covariant derivative is understood in the 
adjoint representation. ${\bf D}\cdot{\bf B}$ and ${\bf D}\cdot{\bf E}$ do not appear, the first because it 
is identically zero after using the Jacobi identity, while the second gives 
vanishing contributions after using the equations of motion.}
\label{tab3}
\begin{tabular*}{\textwidth}{@{}l@{\extracolsep{\fill}}rr}
\hline
\cline{1-3}
        \\
\hline
Gluelumps       & $~$ & $~$ \\
$({\rm Tr \{ O_a} H^a \})$      & $L=1$ & $L=2$ \\ 
$~$ & $~$ & $~$\\\hline
$\Sigma_g^{+\, \prime}$ & ${\bf r}\cdot{\bf E} \;, 
             {\bf r}\cdot({\bf D}\times {\bf B})$  & $~$  \\\hline
$\Sigma_g^-$ & $~$ & $({\bf r}\cdot {\bf D})({\bf r}\cdot {\bf B}) $ \\\hline
$\Pi_g$ & ${\bf r}\times{\bf E}\;, 
             {\bf r}\times({\bf D}\times {\bf B}) $ & $~$ \\\hline
$\Pi_g^{\prime}$ & $~$ & ${\bf r}\times(({\bf r}\cdot{\bf D}) {\bf B} 
              +{\bf D}({\bf r}\cdot{\bf B}))$  \\\hline
$\Delta_g$ & $~$ & $({\bf r}\times {\bf D})^i({\bf r}\times{\bf B})^j 
             +({\bf r}\times{\bf D})^j({\bf r}\times{\bf B})^i$ \\\hline
\hline
$\Sigma_u^{+}$ & $~$ & $({\bf r}\cdot {\bf D})({\bf r}\cdot {\bf E})$\\\hline
$\Sigma_u^-$ & ${\bf r}\cdot{\bf B} \;, 
             {\bf r}\cdot({\bf D}\times {\bf E})$ & $~$ \\\hline 
$\Pi_u$ & ${\bf r}\times{\bf B}\;, 
             {\bf r}\times({\bf D}\times {\bf E})$ & $~$ \\\hline
$\Pi_u^{\prime}$ & $~$ & ${\bf r}\times(({\bf r}\cdot{\bf D}) {\bf E} 
              +{\bf D}({\bf r}\cdot{\bf E})) $ \\\hline
$\Delta_u$ & $~$ & $({\bf r}\times {\bf D})^i({\bf r}\times{\bf E})^j 
             +({\bf r}\times{\bf D})^j({\bf r}\times{\bf E})^i$ \\\hline
\end{tabular*}
\end{table*}
 
So far we have just used the symmetries of pNRQCD at lowest order in the 
multipole expansion. In fact we can go beyond that and predict the shape 
of the static energies by actually studying the correlators 
\begin{eqnarray}
& & \langle 0|H({\bf R},{\bf r}, T/2) H^{\dagger}
({\bf R}^{\prime},{\bf r}^{\prime}, -T/2)|0 \rangle 
\sim \nonumber \\
& & \quad\quad \delta^3({\bf R}-{\bf R}^{\prime})\delta^3({\bf r}-{\bf r}^{\prime})\,e^{-iTV_H(r)}
\end{eqnarray}
for large $T$. At leading order in the multipole expansion we obtain
\begin{eqnarray}
& & \!\!\!\!\!\!\!\!\!\!\! V_H(r) = V_o(r)+ \label{vH}\\
& & \!\!\!\!\!\!\!\!\!+ {i\over T} 
\ln \langle H^a(T/2) \phi(T/2,-T/2)^{\rm adj}_{ab}H^b(-T/2))\rangle,
\nonumber
\end{eqnarray}
where the $T\to\infty$ limit is understood. The general structure of the
gluonic correlator is the following (the contribution from the continuum is
included in the dots)
\begin{eqnarray}
 & & \!\!\!\!\!\!\!\!\!\! \langle H^a(T/2) \phi(T/2,-T/2)^{\rm adj}_{ab}H^b(-T/2)\rangle^{\rm nonpert.} 
\nonumber\\
& &   \!\!\!\!\!\!\! \simeq h \, e^{- i \Lambda_H T} + h^\prime\,e^{- i \Lambda_H^\prime T} + 
\dots  
\label{corglHH}
\,.
\end{eqnarray}  
Since we are in the static 
limit, $1/T \ll \Lambda_{\rm QCD} \sim \Lambda_{\rm H} 
< \Lambda_{\rm H}^{\prime} <  \dots$, one can approximate  the right-hand side of Eq. (\ref{vH}) 
for $T\to\infty$ by just keeping the first exponential of Eq. (\ref{corglHH}). 
Then we get at leading order in the multipole expansion 
\begin{equation}
V_H(r) = V_o(r) + \Lambda_H.
\label{vH2}
\end{equation}
Formula (\ref{vH2}) states that at leading order in the multipole 
expansion the short-distance behavior of the static energies for the gluonic excitations 
between static quarks is described  by the perturbative octet potential plus a nonperturbative constant. 
The constant $\Lambda_H$ depends in general on the particular operator $H$, i.e. on the particular 
gluonic excitation between static quarks. $\Lambda_H$ is the same for operators identifying states 
which are degenerate. 
Notice also that Eq. (\ref{vH2}) can be systematically improved by 
calculating higher orders in the multipole expansion. In particular, 
one can look at how the $O(3)\times$C symmetry is softly broken to $D_{\infty\,h}$ in the short-distance limit.
\begin{figure}[htb]
\vspace{-1cm}
\epsfxsize=7truecm \epsfbox{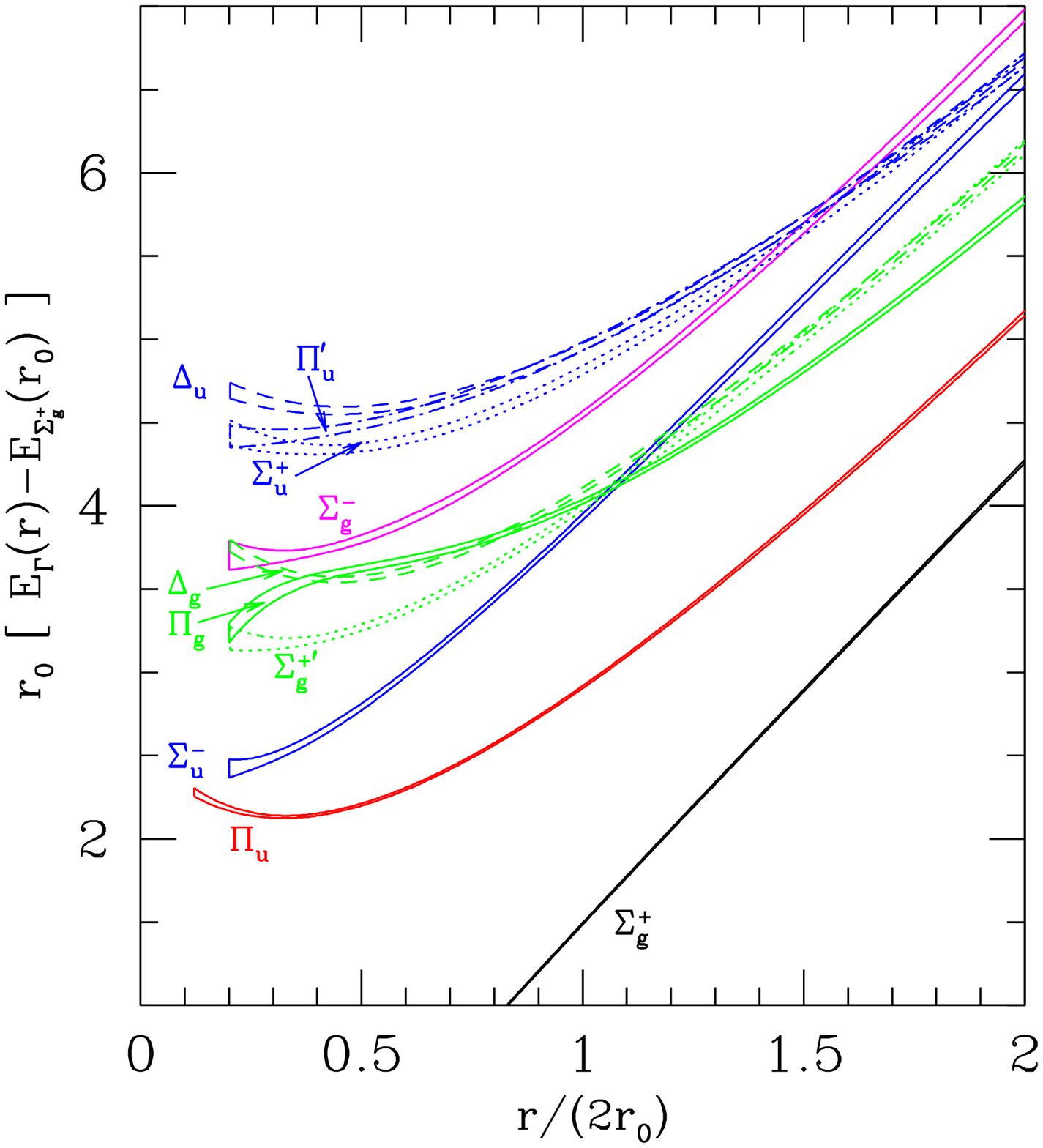}
\vspace{-0.9cm}
\caption{Energies $V_H$ for different gluonic excitation between static quarks at distance $r$  
from the quenched lattice measurements of \cite{Morningstar}, $r_0 \simeq 0.5$ fm. 
The picture is taken from \cite{Michael}.}
\label{fighyb}
\vspace{-0.8cm}
\end{figure}
Let us compare our results against the best available lattice data (see Fig. \ref{fighyb} and \cite{Morningstar}). 
First of all we observe a tendency of all the static energies except $\Pi_g$ to go up at short-distances which may be interpreted as an indication that they want to follow the octet potential shape as required by (\ref{vH2}).   
The $\Pi_u-\Sigma_u^-$ and $\Delta_u-\Pi_u^\prime -\Sigma_u^+$
static energies also show a strong tendency to form a degenerate doublet and triplet respectively at short-distances as 
precisely predicted in (\ref{dege}). However, something strange is
observed for the $g$ static energies. According to (\ref{dege}) a degenerate doublet and triplet should be observed. We can clearly see the doublet $\Pi_g-\Sigma_g^{+\prime}$ but, as mentioned before, the shape of $\Pi_g$ is not compatible with the octet potential. On the other hand $\Delta_g-\Sigma_g^-$ miss 
a state $\Pi_g^{\prime}$ to complete the triplet. We suspect that the plotted
$\Pi_g$ is a superposition of the real $\Pi_g$ in the doublet and the $\Pi_g^{\prime}$ in the triplet. This would explain its peculiar behavior at small $r$.
We expect that future  
lattice simulations will be able to disentangle these two states and confirm our predictions. 
It is interesting to notice that also the hierarchy of the states, 
as displayed in Fig. \ref{fighyb}, is reflected in the dimensionality of the 
operators of Table \ref{tab3}. Lower lying states 
are characterized by lower dimensional operators 
if we use a minimal bases with no electric fields. In particular this allows to understand on simple dimensional grounds the highly non-trivial fact that the doublet and triplet static energies lay close together for the $g$ states but quite far apart for the $u$ ones.

The results given above allow us to relate, in the short-distance limit, the behavior of the 
energies for the gluonic excitation between static quarks with the large time behavior 
of some gluonic correlators, in particular with their correlation 
length. It is particularly appealing that we can extract results for 
the  gauge invariant two-point correlator for the gluon field strength tensor:
$
\langle 0|F^a_{\mu\nu}(t)\phi(t,0)^{\rm adj}_{ab}F^b_{\mu\nu}(0)|0 \rangle  
$,$\phi$ being the adjoint string.
One can parameterize this correlator as a function of two scalar functions:
$\langle 0|{\bf E}^a(t)\phi(t,0)^{\rm adj}_{ab}{\bf E}^b(0)|0 \rangle $
 and $ \langle 0|{\bf B}^a(t)\phi(t,0)^{\rm adj}_{ab}{\bf B}^b(0)|0 \rangle$
with correlations lengths: $T_E=1/\Lambda_{E}$ and $T_B=1/\Lambda_{B}$ respectively. 
From the results of Ref. \cite{Morningstar} displayed in Fig. \ref{fighyb} we can conclude that
\begin{equation}
T_{E} < T_{B}  \label{E>B} \,.
\end{equation}
So far lattice simulations of the gauge invariant two-point correlator for the gluon field strength tensor 
have not reached enough precision to confirm this behavior \cite{latcor}. 
This would be a nice cross-check of lattice simulations. 
On the other hand, recently, a sum rule calculation \cite{Dosch} found evidence 
in favor of Eq. (\ref{E>B}) and also a lattice computation of the 
gluelump masses \cite{Foster} seem to confirm Eq. (\ref{E>B}). 
It would be highly desirable to have more precise lattice data in the short-distance limit in order 
to test our results more quantitatively.

\section{Acknowledgements} I thank A. Pineda, J. Soto and A. Vairo for the  
collaboration which produced  the results contained in the present paper.

\end{document}